\documentclass[twocolumn,aps,prb]{revtex4}
\usepackage{amsmath,amssymb}
\usepackage{graphicx}
\usepackage{epstopdf}
\usepackage[usenames]{color}
\usepackage{amsmath} 
\usepackage{amssymb}	
\usepackage{graphicx} 
\usepackage[usenames]{color}
\normalfont

\renewcommand{\vec}[1]{\ensuremath{\mathbf{#1}}} 
 
\let\baraccent=\= 
\renewcommand{\=}[1]{\stackrel{#1}{=}} 

\begin{document}
\newcommand\bbone{\ensuremath{\mathbbm{1}}}
\newcommand{\ul}{\underline}
\newcommand{\vl}{v_{_L}}
\newcommand{\vc}{\mathbf}
\newcommand{\be}{\begin{equation}}
\newcommand{\ee}{\end{equation}}
\newcommand{\bk}{{{\bf{k}}}}
\newcommand{\bK}{{{\bf{K}}}}
\newcommand{\cE}{{{\cal E}}}
\newcommand{\bQ}{{{\bf{Q}}}}
\newcommand{\br}{{{\bf{r}}}}
\newcommand{\bg}{{{\bf{g}}}}
\newcommand{\bG}{{{\bf{G}}}}
\newcommand{\hbr}{{\hat{\bf{r}}}}
\newcommand{\bR}{{{\bf{R}}}}
\newcommand{\bq}{{\bf{q}}}
\newcommand{\hx}{{\hat{x}}}
\newcommand{\hy}{{\hat{y}}}
\newcommand{\hd}{{\hat{\delta}}}
\newcommand{\bea}{\begin{eqnarray}}
\newcommand{\eea}{\end{eqnarray}}
\newcommand{\ra}{\rangle}
\newcommand{\la}{\langle}
\renewcommand{\tt}{{\tilde{t}}}
\newcommand{\upa}{\uparrow}
\newcommand{\dna}{\downarrow}
\newcommand{\bS}{{\bf S}}
\newcommand{\vS}{\vec{S}}
\newcommand{\dg}{{\dagger}}
\newcommand{\pdg}{{\phantom\dagger}}
\newcommand{\tphi}{{\tilde\phi}}
\newcommand{\cf}{{\cal F}}
\newcommand{\ca}{{\cal A}}
\renewcommand{\ni}{\noindent}
\newcommand{\ct}{{\cal T}}

\title{Competing chiral orders in the topological Haldane-Hubbard model of spin-1/2 fermions and bosons}
\author{C. Hickey$^{1}$, P. Rath$^{1,2}$, and A. Paramekanti$^{1,3}$}
\affiliation{$^1$Department of Physics, University of Toronto, Toronto, Ontario M5S 1A7, Canada}
\affiliation{$^2$Department of Physics, Indian Institute of Technology, Kanpur 208106, India}
\affiliation{$^3$Canadian Institute for Advanced Research, Toronto, Ontario M5G 1Z8, Canada}
\begin{abstract}
Motivated
by experiments on ultracold atoms which have realized
the Haldane model for a Chern insulator, we consider its strongly correlated Mott limit with
spin-$1/2$ fermions. We find that slave rotor mean field theory yields gapped or gapless
chiral spin liquid Mott insulators. To study competing magnetic orders, we consider the strong coupling
effective spin Hamiltonian which includes chiral three-spin exchange. We obtain its classical phase diagram, 
uncovering various chiral magnetic orders including tetrahedral, cone, and noncoplanar spiral
states which can compete with putative chiral quantum spin liquids. We study the effect
of thermal fluctuations on these states, identifying crossovers in the spin
chirality, and phase transitions associated with lattice
symmetry breaking. We also discuss
analogous effective spin Hamiltonians for correlated spin-$1/2$ bosons. Finally, we point out
possible experimental implications of our results for cold atom experiments.
\end{abstract}
\maketitle

\section{Introduction}

Momentum space topology is key to our understanding of phases such as topological
insulators \cite{KaneHasan_RMP2010,QiZhang_RMP2011} or quantum anomalous Hall 
insulators.\cite{HaldaneQAH_PRL1988,Chang_Science2013} The interplay of momentum space
topology and local real space interactions is expected to lead to a rich variety of correlated 
insulating phases.\cite{Pesin2010,Sheng2011,MudryFCI_PRL2011,BernevigFCI_PRX2011,Abanin2012,Cocks2012,
MoessnerFCI_PRL2013,YBKusoc_ARCM2014} This has motivated an extensive investigation of the
time-reversal invariant Kane-Mele-Hubbard model or more realistic variants,\cite{ZhengKMH_PRB2011,HohenadlerKMH_PRL2011,CenkeKMH_PRB2012,HsiangHsuanKMH_PRB2013,HohenadlerKMH_PRB2012,FieteQSHI_PRL2012,LeHur_PRB2013} 
which provide the simplest examples of interacting 
quantum spin Hall insulators. Recently,
cold atoms in a shaken optical lattice \cite{Lewenstein_PRL2012} or laser-induced tunneling or transitions \cite{Jaksch_NJP2003, Lewenstein_JPB2013}
have been proposed for realizing topological states of matter, and employed to realize the 
Haldane honeycomb model of a Chern insulator (CI),\cite{EsslingerHaldane_Nat2014} and Chern bands in the square lattice Hofstadter
model with flux $\pi/2$ per plaquette.
\cite{BlochChern_NPhys2014}
Preliminary experiments suggest that interactions in the Haldane model may not lead to excessive heating.\cite{EsslingerHaldane_Nat2014}
Motivated by this, we
explore strong correlation effects in the Haldane model for spin-$1/2$ fermions and 
bosons.

The Haldane model is defined on the two-dimensional (2D) honeycomb lattice in Fig.~1(a), with a real nearest neighbor hopping 
amplitude $t_1$, and a complex
next-neighbor hopping $t_2 {\rm e}^{i\phi}$ which breaks time-reversal symmetry.\cite{HaldaneQAH_PRL1988} For $t_2=0$, the energy
dispersion is identical to graphene, supporting two inequivalent massless Dirac fermion modes. For small $t_2 \! \neq \! 0$,
and $\phi \neq 0,\pi$, the Dirac fermions acquire a mass, leading to a quantized Hall effect $\sigma_{xy}\!=\! \pm 
e^2/h$ at half-filling. For spin-$1/2$ fermions, with each species at half-filling, $\sigma^{\upa}_{xy}\!=\! \sigma^{\dna}_{xy} \!=\!
\pm e^2/h$. What is the fate of this quantum Hall insulator when interactions lead to Mott localization? Does the Mott insulator support unusual magnetism or spin liquid
ground states?

For spinless fermions, nearest-neighbor repulsion induces a topologically trivial charge density wave insulator, 
\cite{Galitski_PRB2010,Galitski_PRB2011} while spinless bosons with a local Hubbard repulsion form a plaquette Mott
insulator with loop currents.\cite{Hofstetter2014}
A slave-spin approach to the Haldane-Hubbard model suggests the appearance of  fractionalized Chern insulators \cite{Ruegg2013}
or topological N\'eel order
at moderate coupling with a staggered sublattice potential.\cite{Huber2014} 
An alternative slave rotor mean field theory of spin-$1/2$ fermions, in which
Hubbard repulsion localizes only the charge degree of freedom, led to a mean field description of a gapped chiral
spin liquid (CSL).\cite{JingSDW_PRB2011,JingMFT_PRB2011} Gutzwiller projected wave function studies show 
that such a CSL would be in the same phase
as the $\nu\!=\! 1/2$ bosonic Laughlin state.\cite{VishwanathCL_PRB2011,VishwanathCL2_PRB2012}
More generally, such CSLs
have also been shown to be induced by chiral
$3$-spin interactions, \cite{Thomale_PRB2014,BauerKagomeCSL_NComm2014} or
appear naturally for $SU(N)$ fermions  \cite{Hermele_PRL2009} at large $N$.

However, previous work has not considered the various types of magnetic orders which can compete
with quantum spin liquid phases. Indeed, even the
type of short-range spin correlations in the correlated Mott insulator, which could potentially be probed
in cold atom experiments, have not been explored. Finally, since the topological order associated with such chiral
spin liquids will not persist at nonzero temperature, it is important to gain an understanding of the thermal fluctuation
effects on the Mott insulating state if we are to make connection with experiments. However, previous
studies of this model have all been restricted to the $T=0$ limit.

This motivates us to revisit the effect of strong correlations on the Haldane-Hubbard model.
Our key results are the following. (i) We extend the slave rotor mean field theory of the fermionic Haldane-Hubbard model
to larger $t_2$. We show that this parton construction leads to CSLs with gapped or gapless bulk spinon excitations which
descend from CI or Chern metal (CM) phases. The gapless spin liquids display multiple spinon Fermi pockets.
Both types of spin liquids are expected to support gapless spinon edge states.
Realizing the Haldane-Hubbard model with a broad range of $t_2$ would thus allow one to explore the
physics of gapped as well as gapless chiral spin liquid phases.
(ii) The gapped or gapless spin liquids uncovered in the slave rotor parton description
may be unstable to magnetic ordering due to fluctuation effects beyond mean field theory.
To study magnetic instabilities in the strongly correlated limit, we consider the strong coupling limit of the 
Haldane-Hubbard model, and derive and study the effective spin Hamiltonian which has chiral $3$-spin interactions.
This leads us to uncover a rich variety of competing chiral magnetic states such
as tetrahedral, cone, and noncoplanar spirals which could compete with spin liquid phases. Quantum melting
such phases will typically lead to broken symmetry states. However, unlike other states, we find that the tetrahedral 
state has spin correlations which respect all lattice symmetries and a large uniform chirality. We tentatively identify this state
as the classical parent state of a chiral quantum spin liquid. Such tetrahedral states have also been recently
uncovered in a honeycomb lattice Hubbard model at intermediate correlations \cite{ZWang2014}. 
(iii) Turning to physics at nonzero temperature,
we discuss thermal fluctuation effects and lattice symmetry breaking thermal phase transitions associated 
with many of these states using a combination of analytic Landau theory arguments supplemented by
classical Monte Carlo simulations. Such thermal transitions may be observable in experiments.
(iv) Finally, we highlight some analogous results for spin-$1/2$ bosons which can also be studied in cold atom experiments,
but which have not been theoretically explored.

\section{Fermi Haldane-Hubbard model} 
For fermions with spin-$1/2$, the Haldane Hubbard model is described by the 
Hamiltonian
\bea
H_{\rm HH} \!\! &=&\!\! -t_1\! \!\! \sum_{\la i j \ra \sigma} (c^\dg_{i\sigma} c^\pdg_{j\sigma} \!+\! h.c.)
  \!-\!  t_2 \!\!\!\! \sum_{\la\la i j \ra\ra \sigma} \!\! (e^{i\nu_{ij}\phi} 
c^\dg_{i\sigma} c^\pdg _{j\sigma} \!+\! h.c.) \nonumber \\
&+& U \sum_{i} n_{i\uparrow}n_{i\downarrow}
\eea
where $\la.\ra$ and $\la\la.\ra\ra$ denote, respectively, first and second neighbors, and $\nu_{ij}=\pm 1$,
depending on whether we
hop along or opposite to the arrows shown in Fig.~1(a), and $U$ is the local Hubbard repulsion.

\begin{figure}[t]
\includegraphics[scale=0.35]{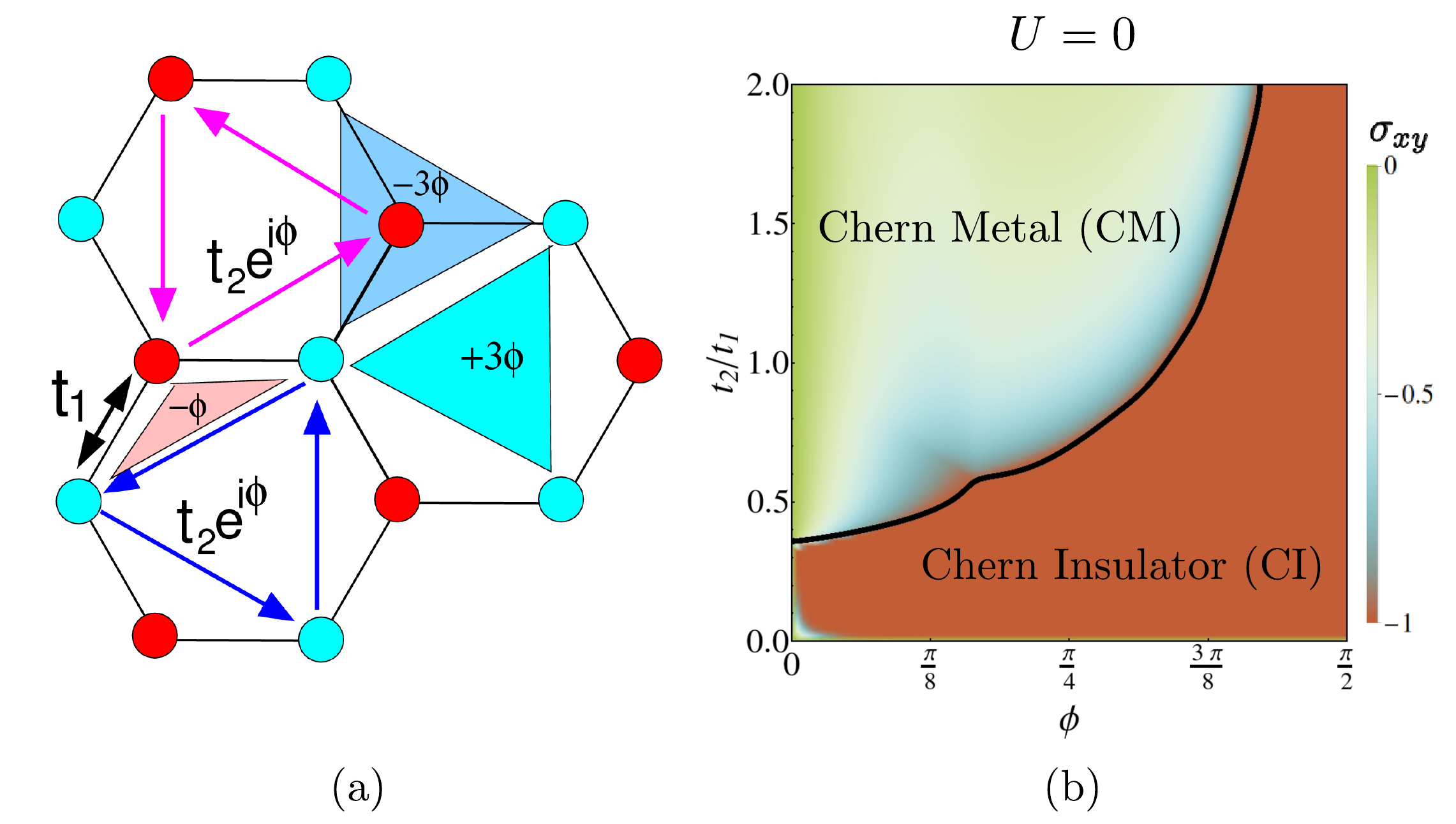}
\caption{(a) Honeycomb lattice showing the two sublattices with hoppings $t_1,t_2$ and phase $\phi$ defining the noninteracting Haldane 
model, with large and small triangles enclosing fluxes $\pm 3\phi$ and $-\phi$ respectively. (b) Phase diagram at half-filling for each spin species 
showing Chern band metal and Chern band insulator, and the Hall conductance per spin $\sigma_{xy}$ in units of $e^2/h$.}
\label{fig:Phase_Diagrams}
\end{figure}

For $U\!=\! 0$, diagonalizing the Hamiltonian leads to Chern bands with
Chern numbers $C\!=\! \pm 1$. Depending on $t_2/t_1$ and $\phi$, these bands may overlap in energy, leading 
to a CM with a non-quantized Hall effect, or be well separated in energy leading to a CI (equivalently, a
quantum Hall insulator) with a Hall conductivity $\sigma_{xy}=\pm e^2/h$ per spin. 
This phase diagram is shown
in Fig.~1(b), along with $\sigma_{xy}$ which is quantized in the CI but non-quantized
in the CM. (Note that the $t_2\! =\! 0$ axis is a Dirac semimetal; on this singular line, the flux plays no role.)

\section{Mean field chiral spin liquids} 
Upon increasing the Hubbard repulsion, this topological band metal or band insulator will transition
into a correlated Mott insulator once interactions exceed the bandwidth. The critical repulsion $U_c \! = \! 8 
\eta | K |$, where $K \! \equiv \! K(t_1,t_2,\phi)$ is the kinetic energy per
site per spin of the noninteracting bands at half-filling. 

Slave rotor mean field theory  \cite{GeorgesSRMFT_PRB2004,ParamekantiSRMFT_PRB2007},
provides a concrete realization of the Mott phase and the Mott transition; it is a parton construction in which the 
spin and charge degrees of freedom are carried by two partons, a neutral spin-$1/2$ fermion and a charged spinless rotor.
The localization of the rotor angular momentum corresponds to the transition into the Mott insulator phase.
A single-site slave rotor mean field theory \cite{GeorgesSRMFT_PRB2004,ParamekantiSRMFT_PRB2007}
yields $\eta \!=\! 1$.
However, a comparison with
quantum Monte Carlo studies of the honeycomb lattice Hubbard model \cite{ZYMengQMChc_Nat2010,Sorella_SRep2012}  at $t_2=0$,
and variational numerical studies
of the triangular lattice Hubbard model \cite{ImadaVarTri_JPSJ2002}  with $t_1\!=\!\phi\!=\! 0$ show that the single
site slave rotor theory overestimates $\eta$; the Mott transition
occurs at a smaller renormalized $\eta \! \approx \! 0.6$-$0.7$.

\begin{figure}[t]
\includegraphics[scale=0.4]{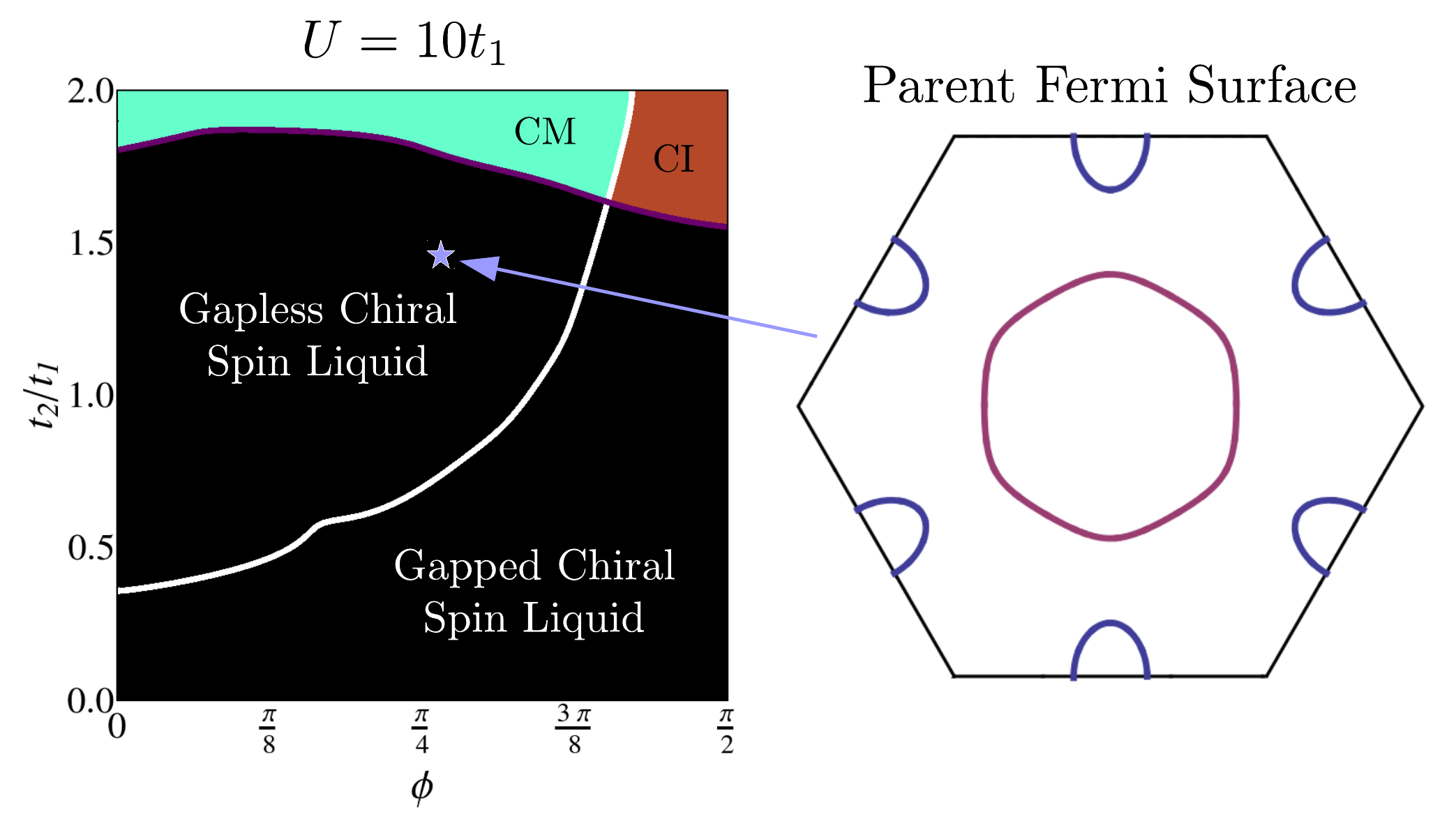}
\caption{Slave rotor mean field theory of the Haldane-Hubbard model showing the topological Chern metal (CM, green),
Chern insulator (CI, brown), and topological spin liquid Mott insulators (black) at $U/t_1=10$. The topological
Mott insulator could have gapless or gapped spinon excitations depending on whether it descends from a CM or
CI phase, leading to gapless or gapped chiral spin liquids. An example of the parent Fermi surface of the 
non-interacting model, which is inherited by the spinons at mean-field level, is shown at a particular point. 
The slave rotor result has been rescaled by
$\eta=0.65$ (see text).}
\label{fig:MottUc}
\end{figure}

Fig.~\ref{fig:MottUc} shows
the phase boundaries which delineate the topological CM or CI from the correlated 
Mott insulator (assuming $U_c \! = \! 8 \eta | K |$ with $\eta\! =\! 0.65$)  at an interaction strength $U/t_1=10$.
Within slave rotor mean field theory, the spinon band structure in the Mott insulator resembles the CI
or CM band structure from which the Mott insulator descends. Thus, depending on $t_2,\phi$, one could have gapped
or gapless spinon excitations in the bulk, accompanied by gapless chiral edge modes; we indicate these mean field
states as `gapped CSL' and `gapless CSL'. The spinon band structure mean field theory is simply inherited from
the parent noninteracting fermion dispersion. Since the
parent Chern metal phase exhibits multiple Fermi pockets, as shown in Fig.~\ref{fig:MottUc}, 
the gapless CSL will inherit these as spinon Fermi pockets. Gapless chiral spin liquids have also been recently 
discussed in kagome antiferromagnets.\cite{Bieri2014}

Going beyond mean field 
theory, we expect the `gapped CSL' to become a genuine spin liquid with semion
excitations and topological degeneracy.\cite{VishwanathCL_PRB2011,VishwanathCL2_PRB2012}  While these CSLs
are expected to be stable for weak gauge fluctuations,
strong gauge fluctuations might drive magnetic ordering instablilties, or, spinon pairing instabilities in the case of the 
gapless CSL.\cite{Laughlin1988,Nayak2007}
Since slave rotor theory does not capture competing magnetic ordering tendencies of the Mott insulator,
we address such potential competing phases using a strong coupling
expansion.\cite{MacDonaldtU_PRB1988}

\section{Competing magnetic orders} At large $U$, a standard derivation \cite{MacDonaldtU_PRB1988,Motrunich_PRB2006}
leads to an effective spin Hamiltonian
\bea
H_{\rm spin} \!\!\!&=&\!\!\! \frac{4 t_1^2}{U} \sum_{\la ij \ra} \vec S_i \cdot \vec S_j
 + \frac{4t_2^2}{U}\sum_{\la\la ij \ra\ra} \!\! \vec S_i \cdot \vec S_j \nonumber \\
\!\!\!&-&\!\!\! \frac{24t_1^{2}t_2}{U^2} \!\!\!\!\! \sum_{{\rm small}-\triangle} \!\!\!  \hat{\chi}_\triangle \sin\Phi_\triangle
- \frac{24t_2^{3}}{U^2} \!\!\!\! \sum_{{\rm big}-\triangle} \! \hat{\chi}_\triangle \sin\Phi_\triangle
\label{Hspinfermion}
\eea
where $\hat{\chi}_\triangle \equiv \vec S_i \cdot \vec S_j \times \vec S_k$ is the scalar spin chirality operator defined
around triangular plaquettes, with the sites $\{ijk\}$ being labelled going anticlockwise around the triangles. 
$\Phi_\triangle$ being the flux enclosed by the corresponding triangle.
As shown in Fig.~1(a), the set of triangles includes small triangles that enclose flux $\Phi_\triangle= - \phi$, big
triangles within a hexagon which enclose a flux $\Phi_\triangle= 3 \phi$, and big triangles defined around a
site which enclose a flux $\Phi_\triangle= -3 \phi$.

Treating the spin as a classical vector,
we have obtained the magnetically ordered ground states 
of $H_{\rm spin}$ using numerical simulated annealing and variational spin
configurations. This approximation may be formally set up by scaling the $2$-spin exchange couplings by $1/(4 S^2)$ 
and the $3$-spin exchange couplings by $1/(8 S^3)$ and then taking $S \to \infty$ which leads to
a classical spin model with quantum fluctuations suppressed by ${\cal O}(1/S)$. Below we will use this classical 
approximation for $S=1/2$
as an uncontrolled approximation, which is nevertheless known to work well in many cases in predicting the correct
magnetic order in 2D quantum spin models.
For plotting the ground state spin configurations and structure factors, we view the honeycomb lattice as a 
square lattice with $1/4$ deleted bonds (`brickwall' lattice), similar to the geometry
realized in optical lattice experiments.\cite{EsslingerHaldane_Nat2014}
The resulting phase diagram,
shown in Fig.~3 for $U/t=10$, contains six phases: N\'eel, Tetrahedral, Cantellated Tetrahedral, Spiral, Cone-I and Cone-II,
with a line indicating the regime below which this strong coupling description is appropriate.

Our main finding, atleast for $t_2/t_1 \lesssim 1.5$, as discussed below, is that the gapped spin liquid phase found in the slave rotor mean field theory 
in Section III may be
potentially unstable to {\it commensurate} magnetic orders such as
Neel or Tetrahedral order. On the other hand, the regime we identify as a gapless spin liquid with Fermi pockets
in the slave rotor mean field theory appears to be dominantly unstable to various {\it incommensurate} spiral phases 
such as Spiral or Cone phases. 
Since the phase diagrams obtained in Fig.~2 and Fig.~3 involve different approximations, neither one is fully
reliable; further numerical work is necessary to definitely identify whether the spin liquid
phases are indeed stable or give way to magnetic or other symmetry breaking orders. However, the true phase
diagram is expected to display phases shown in the phase diagrams in Fig.~2 and Fig.~3.

We describe these various competing magnetic ground states below. In the Appendix, we include
``Common origin'' plots \cite{Henley_PRB2013} for these states for further insights.

\ni (i) \underline{\it N\'eel}: The N\'eel state is collinear with spins pointing opposite to each other on the two 
 sublattices of the brickwall lattice, with $\la \hat{\chi}_\triangle \ra=0$ on all triangular plaquettes. Its
 structure factor ${\cal S}(\bq)= \frac{1}{N} \sum_{i,j} \la \vec S_i \cdot \vec S_j \ra 
 {\rm e}^{i\bq\cdot(\vec r_i - \vec r_j)}$ exhibits a Bragg peak at $\bq=(\pi,\pi)$. 
 
\begin{figure}[t]
\includegraphics[scale=0.25]{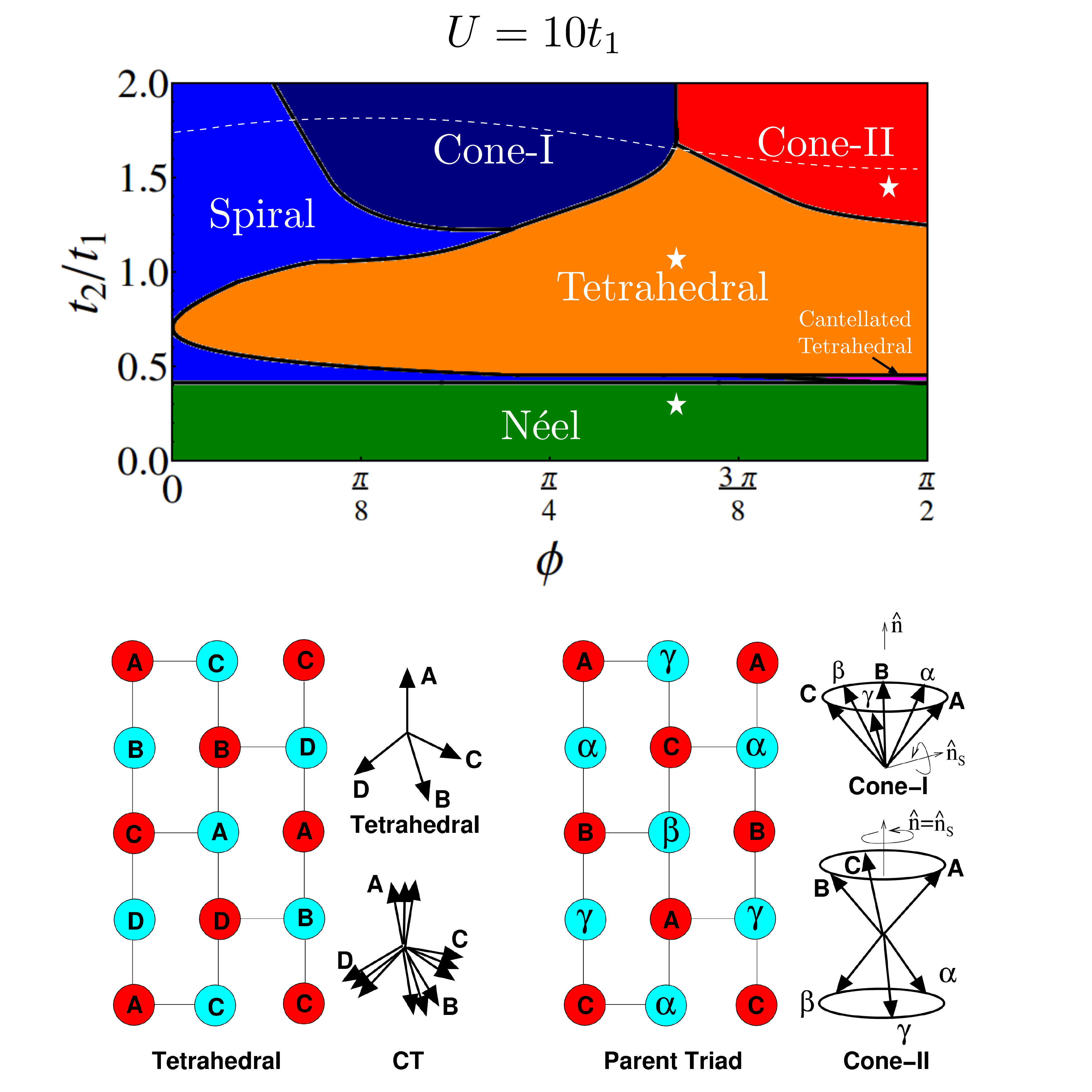}
\caption{\underline{Top:} Phase diagram of $H_{\rm spin}$ showing various classical magnetically ordered ground states at $U/t=10$:
N\'eel, Spiral, Tetrahedral, Cantellated Tetrahedral (CT), Cone-I, and Cone-II. Below the dashed (white) line, the slave rotor result
shows that we are in the Mott insulator, so the strong coupling expansion is expected to be valid.
Stars indicate points where we numerically study the effect of
thermal fluctuations.
\underline{Bottom:} Spin configurations in the Tetrahedral and
Cone states (viewing the honeycomb lattice as a `brickwall' for visual convenience). The indicated Triad states act as
parent states for Cone-I and Cone-II, which are obtained by spiralling the Triads about the indicated
$\hat{n}_s$ axis. The CT descends from the Tetrahedral by a $\sqrt{3}\times\sqrt{3}$ canting pattern.}
\label{fig:SpinPhases}
\end{figure}

\ni (ii) \underline{\it Tetrahedral}: In this state, the spins point from the origin towards the four corners of a tetrahedron, and
are tiled on the lattice as shown in Fig.~\ref{fig:SpinPhases}. This noncoplanar state has 
a uniform chirality $\la \hat{\chi}_\triangle \ra= - 1/6\sqrt{3}$ on each small-$\triangle$.
The tetrahedral state is a triple-$\bq$ state, with the structure factor ${\cal S}(\bq)$ exhibiting 
Bragg peaks at $\bq=\pm (\pi/2,\pm \pi/2)$ and $\bq=(0,\pi)$.

\ni (iii) \underline{\it Cantellated Tetrahedral}: The Cantellated Tetrahedral (CT) state descends from a
parent tetrahedral state. It is reminiscent
of the cuboctahedral state in frustrated kagome antiferromagnets. \cite{Lhuillier2011,Ghosh2014,Sheng2014}
In the parent tetrahedral configuration, the spins marked $A$ (or $B,C,D$) 
in Fig.~\ref{fig:SpinPhases} form a honeycomb (or brickwall) lattice, with a 
larger unit cell. We can deform these ferromagnetically aligned spins by dividing the larger honeycomb lattice into 
two sublattices, and then allowing for a $\sqrt{3}\times\sqrt{3}$ canting on each sublattice; this splits each
vertex of the tetrahedron into six vertices (forming a small hexagon) leading to the CT state with a $24$-site unit cell. 
Here, the CT state exists over an extremely narrow window between the N\'eel and Tetrahedral states.
Third-neighbor $\vec S_i \cdot \vec S_j$ terms, which favor the cuboctahedral
state on the kagome lattice, could enhance the regime of CT order.

(iv) \underline{\it Cone-I}: This state descends from a Triad-I state in which spins on the large-$\triangle$ tend to form a 
triad on each sublattice, leading to a nonzero chirality. The organization of the spins is shown in
Fig.~\ref{fig:SpinPhases}. In Cone-I, which occurs for $\phi < \pi/3$, we start with triads on the two sublattices that
are lined up with a common axis $\hat{n}$, with spins making an angle $\theta$ with $\hat{n}$. Forming an
orthonormal basis $\{\hat{n},\hat{n}_{\perp 1},\hat{n}_{\perp 2}\}$, we parameterize the three 
spins of the triad on sublattice-A, $\vec S^A_j$ ($j \!=\! 0,1,2$), as $\vec S^A_j \!=\! \cos\theta\hat{n} \!+\! 
\sin\theta\cos(2\pi j/3)\hat{n}_{\perp 1} \!+\! \sin\theta\sin(2\pi j/3) \hat{n}_{\perp 2}$, while the triad on 
sublattice-B is rotated by $\varphi$ about $\hat{n}$, with $\vec S^B_j = \cos\theta\hat{n} + 
\sin\theta\cos(\varphi+2\pi j/3)\hat{n}_{\perp 1} + \sin\theta\sin(\varphi+2\pi j/3) \hat{n}_{\perp 2}$. This state has a 
nonzero net magnetization. The ground state energy is independent of $\varphi$.
The triad angle $\theta$ varies with $t_2,\phi$;
at large $t_2/t_1$, the spins on each large-$\triangle$ tend to form an
orthonormal triad, with $\theta=\tan^{-1}(1/\sqrt{2})$.
The Triad-I is a triple-$\bq$ state; ${\cal S}(\bq)$ exhibits Bragg peaks at
$\bq=(0,\pm 2\pi/3)$, $\bq=(\pi,\pm\pi/3)$, and  $\bq=(0,0)$.
Let us now consider
an axis $\hat{n}_s$ which is perpendicular to the plane formed by $\hat{n}$ and any one of the spins. The Cone-I state 
is obtained by rotating all the spins  about this axis by an angle $\theta(\br)=\bq \cdot \br$ where $\bq$ is 
an incommensurate spiral wavevector; moving along $\bq$, 
each spin of the triad again traces out a cone, and the Bragg peaks acquire a weak incommensuration.

(v) \underline{\it Cone-II}: The Cone-II state occurs for $\phi>\pi/3$, where the $\sin 3\phi $ term in Eq.~\ref{Hspinfermion}
changes sign. Here, the parent state is a Triad-II state in which
the spins on sublattice-A are reordered, with $\vec S^A_j \!=\! \cos\theta\hat{n} + 
\sin\theta\cos(4\pi j/3)\hat{n}_{\perp 1} + \sin\theta\sin(4\pi j/3) \hat{n}_{\perp 2}$, while the cone on sublattice-B 
is simply flipped, so that 
$\vec S^B_j = - \cos\theta\hat{n} + \sin\theta\cos(\varphi+2\pi j/3)\hat{n}_{\perp 1} + \sin\theta\sin(\varphi+2\pi j/3) \hat{n}_{\perp 2}$,
with the ground state energy being independent of $\varphi$.
The Triad-II state has a nonzero staggered magnetization, with Bragg peaks in ${\cal S}(\bq)$ at $\bq=(0,\pm 2\pi/3)$, $\bq=(\pi,\pm\pi/3)$, and  
$\bq=\pm(\pi,\pi)$. The Cone-II state 
is obtained by rotating all the spins of Triad-II about the $\hat{n}_s = \hat{n}$ axis by $\theta(\br)=\bq \cdot \br$ where $\bq$ is 
an incommensurate spiral wavevector, leading to weakly incommensurate Bragg peaks.

\ni (vi) \underline{\it Noncoplanar Spiral}: At $\phi=0$, the Hamiltonian $H_{\rm spin}$ is the frustrated $J_1$-$J_2$ Heisenberg
antiferromagnet on the honeycomb lattice. The classical ground states of this model 
consists of coplanar incommensurate spirals for $J_2 > J_1/6$. \cite{ParamekantiJ1J2_PRB2010} For $\phi \neq 0$, chiral terms in
$H_{\rm spin}$ cause spins on the $A$ and $B$ sublattices to
cant in opposite directions away from the spiral plane, leading to a noncoplanar spiral with a 
uniform $\chi_\triangle \neq 0$ on all big-$\triangle$s.

\section{Thermal Fluctuations} 
In 2D, the Mermin-Wagner theorem  \cite{MerminWagner_PRL1966}
precludes spin $SU(2)$ symmetry breaking at $T>0$. However, discrete orders associated with
lattice symmetry breaking can survive thermal fluctuations.

\begin{figure}[t]
\includegraphics[scale=0.2]{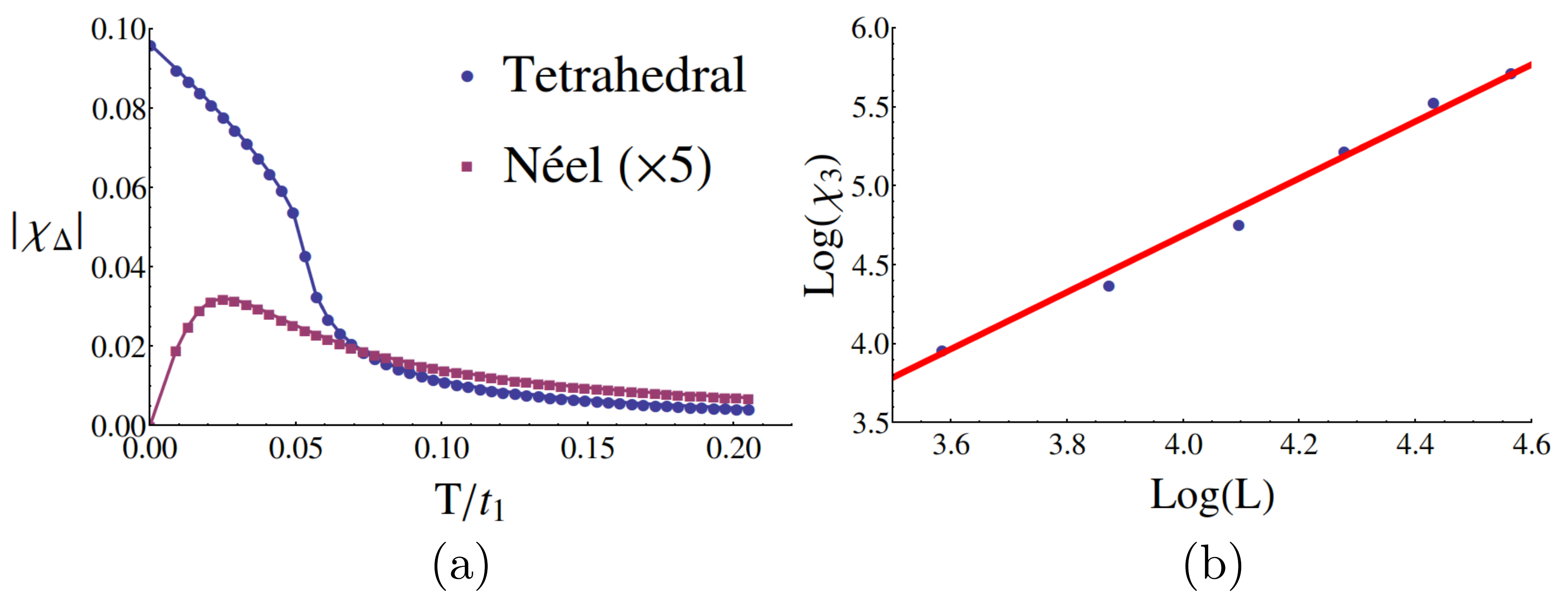}
\caption{(a) Temperature dependence of the spin chirality $\chi_\triangle$ on small triangles for $U/t\!=\! 10$ and $\phi\! =\! \pi/3$, 
showing its low temperature saturation in the Tetrahedral phase ($t_2/t_1\! =\! 1$) and its vanishing in the $T\! =\! 0$
N\'eel phase ($t_2/t_1 \! =\! 0.4$). We have plotted $5 \times \chi_\triangle$ in the N\'eel state for clarity. (b) Log plot of finite size scaling of 
critical susceptibility for the $Z_3$-clock ordering transition in the
Cone-II state at $U/t_1\!=\! 10$, $\phi\! =\! 9\pi/20$, $t_2/t_1\!=\! 1.5$ showing $\chi_3 \!\sim \! L^{\gamma/\nu}$, with $\gamma/\nu \!=\! 1.80(8)$.}
\label{fig:finiteT}
\end{figure}

The N\'eel and 
Tetrahedral orders break $SU(2)$ symmetry, but their
$SU(2)$ invariant spin correlations, such as $\la \vec S_i \cdot \vec S_j\ra$ and 
$\la \vec S_i \cdot \vec S_j \times \vec S_k \ra$, respect all lattice symmetries. These states at $T>0$
can thus be smoothly connected to the high-$T$ paramagnetic state. As shown in Fig.~\ref{fig:finiteT}(a),
for the Tetrahedral state at $U/t_1=10$,
there is a crossover temperature $\sim 0.05 t_1$ below which the spin chirality $\chi_\triangle$ on 
small triangles
becomes large, saturating to $1/6\sqrt{3}$ as $T \to 0$. The $T>0$ regime
here can thus be viewed as a {\it classical CSL}. 
By contrast, $\chi_\triangle$ vanishes at $T=0$ in the N\'eel state, but it exhibits a peak at nonzero temperature.

Previous work has shown that at $\phi=0$, the spiral ground states of the frustrated $J_1$-$J_2$ Heisenberg
antiferromagnet have nematic order associated with 
broken $C_3$ honeycomb lattice symmetry which survives up to a thermal $Z_3$-clock transition. \cite{ParamekantiJ1J2_PRB2010} The
noncoplanar spiral does not involve any additional symmetry breaking, so it undergoes a similar $Z_3$-clock
transition at $\phi \neq 0$.

In the Cone states, the choice of the common axis $\hat{n}$ represents a spontaneously
broken $SU(2)$ symmetry; long wavelength fluctuations will disorder $\hat{n}$ at arbitrarily
small $T >0$, restoring $SU(2)$ symmetry. 
Using Monte Carlo simulations, we find that the nearest neighbor spin correlations $\la \vec S_i \cdot \vec S_j \ra$
are modulated in the Cone states, leading to energy modulations on the bonds of the lattice. In the
Cone-I state, these energy modulations
resemble a
$\sqrt{3}\times\sqrt{3}$ columnar valence-bond solid pattern, while
in the Cone-II state the energy modulations on the bonds resemble a staggered
valence-bond solid bond pattern. This leads to a $Z_3$ symmetry breaking in
both states.
We thus expect the Cone states to undergo a thermal $Z_3$-clock transition into the 
high temperature paramagnetic state. In the Cone-II state at $U/t_1\!=\! 10$, $\phi\!=\! 9\pi/20$, $t_2/t_1\!=\!  1.5$, 
the computed peak susceptibility $\chi_3(L)$ for this transition is plotted in Fig.~\ref{fig:finiteT}(b); 
its finite size scaling shows $T_c \approx 0.10 t_1$, with $\chi_3(L) \!\sim\! L^{\gamma/\nu}$, with $\gamma/\nu\! =\! 1.80(8)$, consistent with
the exact $Z_3$-clock result  \cite{WuPotts_RMP1982} for the exponent ratio $\gamma/\nu= 26/15$.
 
\section{Mott insulator of spinor bosons in the Haldane Hubbard model}
Cold atom experiments can also study interacting bosons in the Haldane-Hubbard model. Previous work on
spinless bosons has shown the emergence of chiral superfluids 
and plaquette Mott insulators with loop currents. \cite{Hofstetter2014}
For pseudospin-$1/2$ bosons, the Haldane-Hubbard model is
\bea
H^{\rm Bose}_{\rm HH} \!\! &=&\!\! -t_1\! \!\! \sum_{\la i j \ra \sigma} (b^\dg_{i\sigma} b^\pdg_{j\sigma} \!+\! h.c.)
  \!-\!  t_2 \!\!\!\! \sum_{\la\la i j \ra\ra \sigma} \!\! (e^{i\nu_{ij}\phi} 
b^\dg_{i\sigma} b^\pdg _{j\sigma} \!+\! h.c.) \nonumber \\
&+& \frac{1}{2} \sum_{i\sigma\sigma'} U_{\sigma\sigma'} n_{i\sigma}n_{i\sigma'}
\eea
For atoms such as $^{87}$Rb, the background scattering
lengths are nearly isotropic, so we set $U_{\sigma\sigma'}=U$.

Here, we focus on the strong coupling
limit $U/t_1,U/t_2 \gg 1$. The spin Hamiltonian in the resulting Mott insulator can be
derived in a manner similar to the fermion case, although it is
more tedious to calculate the terms to ${\cal O}(t^3/U^2)$. We find
\bea
H^{\rm Bose}_{\rm spin} \!\!\!&=&\!\!\! J_1(\phi) \sum_{\la ij \ra} \vec S_i \cdot \vec S_j
 + J_2(\phi) \sum_{\la\la ij \ra\ra} \!\! \vec S_i \cdot \vec S_j \nonumber \\
 &+&\!\!\! \frac{24t_1^2 t_2}{U^2}
 \!\!\!\! \sum_{{\rm small}-\triangle} \!\!\! \hat{\chi}_\triangle  \sin \Phi_\triangle
+ \frac{24 t_2^3}{U^2} \! \sum_{{\rm big}-\triangle} \! \hat{\chi}_\triangle \sin \Phi_\triangle
\eea
with
$J_1(\phi) \!\!=\! -\! 4\frac{t_1^2}{U} \! -\! 24 \frac{t_1^2 t_2}{U^2} \cos \phi$
and $J_2(\phi) \! = \! - \frac{4t_2^2}{U} \!-\! 6 \frac{t_1^2 t_2}{U^2}\cos\phi \!-\! 12 \frac{t_{2}^3}{U^2} \cos3\phi$.
Similar to the fermion case, the set of triangles includes small triangles that enclose flux $\Phi_\triangle= - \phi$, big
triangles within a hexagon which enclose a flux $\Phi_\triangle= 3 \phi$, and big triangles defined around a
site which enclose a flux $\Phi_\triangle= -3 \phi$.
As expected, the two-spin exchanges are ferromagnetic at $\phi\!=\! 0$, 
and the chiral terms also change sign relative to the fermions.

In order to localize bosons, it is well known that the required repulsive interactions are much
stronger. Thus, $t/U$ is much smaller in the Mott insulating phase of spinor bosons, so that
the chiral terms are much weaker
in magnitude compared to the fermion case. In addition, unlike the fermion case, both $J_1$
and $J_2$ are ferromagnetic, which means these two-spin exchange couplings do not frustrate 
each other. Due to both these reasons, we find that ferromagnetic order almost completely dominates
the magnetic phase diagram in the Mott insulator of spinor bosons.  

In the ferromagnetic Mott insulator, the spin of the bosons plays no role. Thus, our model should yield
results identical to that for spinless bosons. Indeed, assuming ferromagnetic order in our model, the
chiral terms vanish. Nevertheless, we
find that the bond energies displays a flux dependence arising from the flux-dependence of the
two-spin exchange couplings. This leads to a weak nonzero loop currents $\sim t^3/U^2$ around triangular
plaquettes, 
consistent with weak loop currents predicted in the plaquette Mott
insulator of spinless bosons using quite different approaches. \cite{Hofstetter2014} This result for spinor bosons 
is in striking contrast to the fermion case,
where the only flux dependence arises from the chiral terms, so loop currents to ${\cal O}(t^3/U^2)$ only
arise in fermion Mott insulators which have non-coplanar spins with a nonzero scalar chirality.

\section{Discussion} 
We have discussed spin-$1/2$ fermions or bosons in the strongly correlated Haldane-Hubbard
model. Our main finding is the rich variety of competing chiral magnetic orders 
with varying flux and hopping. Such orders, even if they are short-ranged, could be probed 
via magnetic Bragg scattering,\cite{Corcovilos2010,Hart2014} which has been shown to be a useful probe in
atomic gases. For fermions, since $\partial H_{\rm spin}/\partial\phi \sim \hat{\chi}_\triangle$, measuring 
the energy change upon an adiabatic change of flux is an indirect way to measure $\chi_\triangle$.
This relation also ties the spin chirality to (weak) charge currents in the Mott insulator which could
potentially be detected using quantum quenches. \cite{Killi_PRA2012a,Killi_PRA2012b,Dhar_PRB2013}
More direct routes to detect the spin chirality of atoms are desirable.
Doping these magnetically ordered states leads to 
nontrivial Chern bands and a nonzero charge Hall effect, similar to that in certain
frustrated metallic magnets. \cite{Martin2008,Martin2010,Kumar2012}
Quantum melting the Tetrahedral state, which is uniform with a large 
chirality, may lead to a quantum CSL ground state with short range magnetic correlations at the same 
wavevectors. Further work is needed to clarify the energetic competition between CSLs and chiral magnetic orders.
Finally, although the Mott insulating phase of spinor bosons is dominated by the ferromagnetic phase,
lowering $U$ might lead to superfluids with exotic magnetic orders; this will be explored elsewhere.

\acknowledgments

We acknowledge useful discussions with I. Bloch,
D. Greif, Z. Papic, S. Parameswaran, and N. Perkins. This research was funded by NSERC
of Canada and Mitacs through the Mitacs-Globalink program. AP thanks the
Aspen Center for Physics (Grant No. NSF PHY-1066293) where part of this work was completed.

\widetext

\appendix

\section{Common origin plots for the various magnetic orders}

For visualizing spin configurations, we employ common origin plots \cite{Henley_PRB2013} where all spins on the lattice are plotted with their 
tails at the center of a unit sphere and heads marked as dots on the surface of the sphere. We mark the spins on the two sublattices of
the honeycomb (i.e., brickwall) lattice using different colors, red and black. Doing this for the Tetrahedral state, for instance, we see that all
spins point along one of four directions, towards the vertices of a tetrahedron. For the Cantellated Tetrahedral, the vertices split into six points, 
three on each sublattice. For the Spiral state, we find that the spins on the two sublattices which support incommensurate spirals
cant out of the plane of the coplanar spiral in opposite directions. The Cone-I state, which is obtained by spinning a triad state about a
specific axis leads to one great circle and two small circles, while Cone-II leads to two parallel small circles.
\begin{figure}[tb]
\includegraphics[scale=0.2]{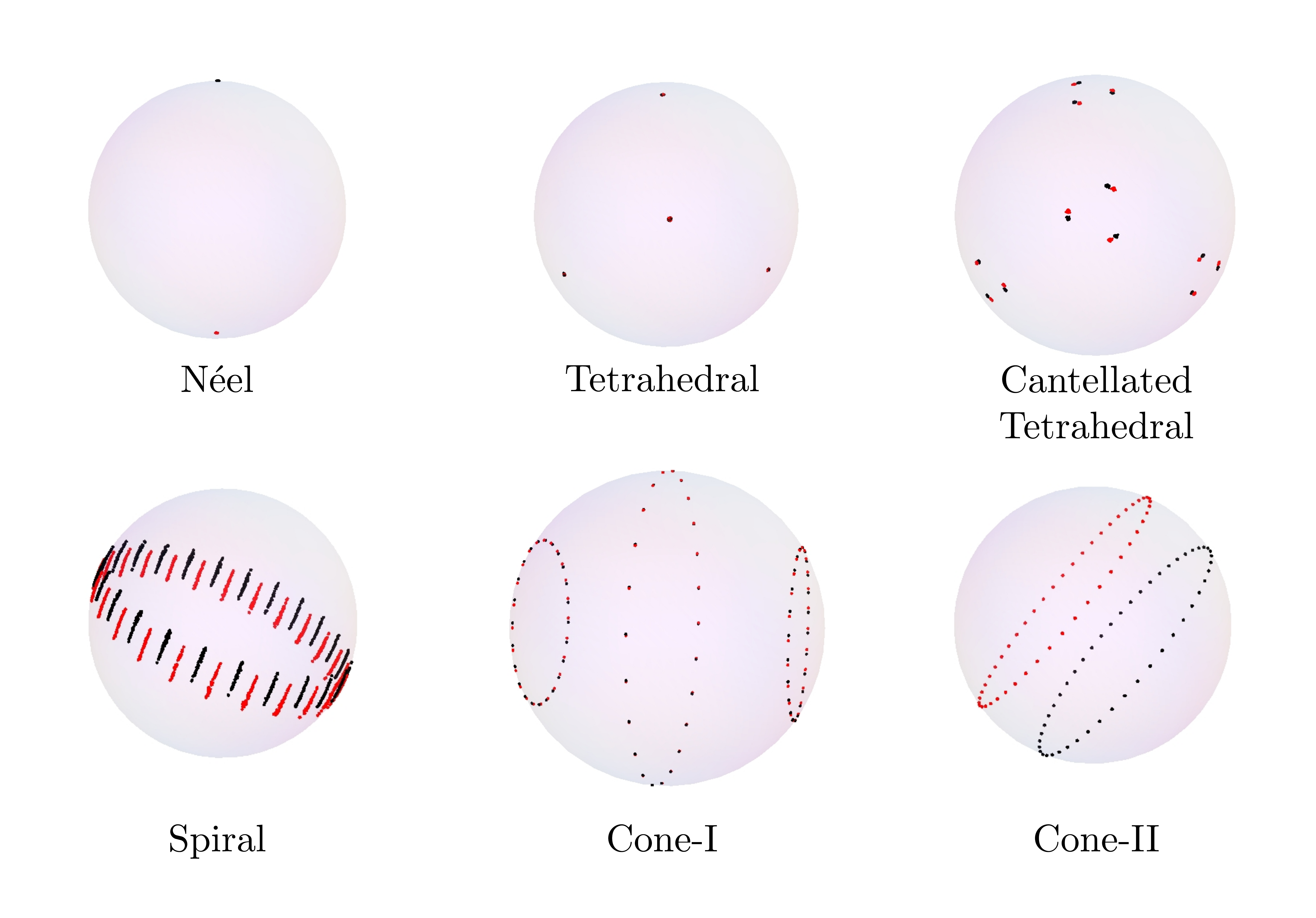}
\label{fig1}
\caption{Common origin plots for the Neel, Tetrahedral, Cantellated Tetrahedral, Spiral, Cone-I and Cone-II states, with red and black dots representing spins on the
two sublattices of the honeycomb (or `brickwall') lattice.}
\end{figure}

\end{document}